# A Model for Athermal Strain Localization in Dry Sheared Fault Gouge:

Xiao Ma

Department of Civil and Environmental Engineering,

University of Illinois at Urbana-Champaign,

Urbana, IL61801, USA

xiaoma5@illinois.edu

A. E. Elbanna

Department of Civil and Environmental Engineering,

University of Illinois at Urbana-Champaign,

Urbana, IL61801, USA

elbanna2@illinois.edu

## Key Points:

1. We propos a physics-based numerical model for strain localization in the fault gouge.
2. Complex shear band localization pattern matches the field and experiment observations.
3. Identify a number of brittle to ductile transition mechanisms.

# Abstract:

Shear banding is widely observed in natural fault zones as well as in gouge layers in laboratory experiments. Understanding the dynamics of strain localization under different loading conditions is essential for quantifying strength evolution of fault gouge, energy partitioning during earthquakes and characterizing rheological transitions and fault zone structure changes. To that end, we develop a physics-based continuum model for strain localization in sheared granular materials. The grain-scale dynamics is described by the Shear Transformation Zone (STZ) theory, a non-equilibrium statistical thermodynamic framework for viscoplastic deformation in amorphous materials. Using a finite strain computational framework, we investigate the initiation and growth of complex shear bands under a variety of loading conditions and identify implication for strength evolution and ductile to brittle transition. Our numerical results show similar localization patterns to field and lab observations and suggest that shear zones show more ductile response at higher confining pressures, lower dilatancy and loose initial conditions. Lower pressures, higher loading rates and higher dilatancy favor a brittle response and larger strength drops. These findings shed light on a range of mechanisms for strength evolution in dry sheared gouge and provide a critical input to physics-based multiscale models of fault zone instabilities.

# 1. Introduction

Fault gouge is a highly-granulated material which is formed in many fault zones due to the fragmentation of the intact country rocks as the fault develops under progressive shearing. Studies have shown extreme localization of slip within gouges layers that in some cases may be less than a few millimeters thick [Chester and Chester, 1998; De Paola et al., 2008; Boullier et al., 2009; Heesakkers et al., 2011; Smith et al., 2011; Fondriest et al., 2012]. Origin and evolution dynamics of this shear banding at lab and field scales are not yet fully understood.

Strain localization in granular material has been identified and studied extensively, mostly at low strain rates, using the tri-axial apparatus and direct double-shear test machines [Logan et al., 1979; Marone et al., 1990; Beeler et al., 1996; Marone, 1998; Logan, 2007; Rathbun and Marone, 2010; ]. Localization has been cited as a mechanism for material weakening in fault zones and has been linked to strain softening and changes in frictional rate sensitivity [Marone, 1998]. These rheological changes play a critical role in determining the stability of sliding during nucleation and subsequent propagation of earthquakes and control a variety of source parameters including strength drop, slip weakening distance and radiation efficiency [Marone, et al. 1990; Kanamori and Heaton, 2000; Rice, 2006; Logan, 2007]. Understanding the interplay between strain localization dynamics and fault zone constitutive response is thus crucial for resolving several outstanding challenges in earthquake physics.

Strain localization in sheared fault gouge has also been studied numerically [Platt et al.,2014, Rice et al., 2014]. The focus was primarily on thermal mechanisms for strain localization including pore fluid thermal pressurization and chemical decomposition of carbonate rich gouge due to shear heating. These models were limited to idealized 1D geometry in the direction perpendicular to the mathematical fault plane. Daub and Carlson [2009] and Hermundstadt et al., [2010] investigated athermal shear localization in gouge due to inhomogeneous plastic deformation. Lieou et al. [2014] and Kothari and Elbanna [2016] have extended this framework to investigate the influence of grain fragmentation and acoustic vibrations on strain localization dynamics respectively. These models were also limited to 1D idealization of fault zones. Field and lab observations suggest that shear bands have complicated structure including Riedel, boundary and Y-shears. Such complexity is beyond the capacity of 1d models. Here, we present a continuum 2D model for strain localization in visoplastic gouge layers to capture the evolution of this complex texture.

Our primary theoretical tool for investigating fault zone inelasticity is the Shear Transformation Zone (STZ) framework [Falk and Langer 1998]. The shear transformation zone (STZ) theory is a continuum model of plastic deformation in amorphous solids that quantifies local configurational disorder [Falk and Langer 1998]. The basic assumption in the theory is that plastic deformation occurs at rare non-interacting localized spots known as shear transformation zones (STZs). An internal state variable, the effective temperature, or compactivity [Falk and Langer, 2011; Lieou et al. 2014], describes fluctuations in the configurational states of the granular material (i.e., a measure of local entropy) and controls the density of STZs [Lieou et al. 2014]. Effective temperature is related to the system porosity [Lieou et al. 2014]. This approach coarse-grains granular simulations while retaining important physical concepts. The STZ framework has been recently extended to resolve additional several gouge-specific mechanisms, such as grain breakage [Lieou et al. 2014], and compaction under vibration [Lieou et al, 2016, Kothari and Elbanna, 2016] beyond what is possible using classical rate and state formulations [Dieterich 1979; Ruina, 1983]. Elbanna and Carlson [2014] have also extended the flash heating theory [Rice, 2006] to fault gouge within the STZ framework.

The new contribution of this paper is the implementation of the STZ theory in a multi-dimension finite element framework enabling us to capture the initiation of the shear bands, the evolution of shear localization patterns, and changes in gouge strength with slip. Furthermore, it enables exploring the effect of spatial heterogeneities on fault zone structure and strength beyond what is possible in idealized 1D models.

The remainder of the paper is organized as follows. In Section 2 we review the basic elements of the STZ theory. In Section 3 we discuss the implementation of the STZ theory into a finite strain visco-plastic finite element framework. In Section 4, we present our numerical results for the evolution of shear bands, and factors affecting the shear band localization. In Section 5, we discuss the implications of our findings for fault zone dynamics and multiscale modeling of earthquakes. We summarize our conclusions in Section 6.

## 2. Review of Shear Transformation (STZ) theory

STZ theory is a nonequilibrium statistical thermodynamic framework for describing plastic deformations in amorphous materials by quantifying local disorder. It has been applied to a variety of systems including granular fault gouge [Daub and Carlson, 2008; Daub et al., 2008; Daub and Carlson, 2010; Hermundstad et al., 2010], glassy materials [Falk and Langer, 1998, 2000; Manning et al., 2007, 2009], thin film lubricants [Lemaitre and Carlson, 2004], and hard spheres [Lieou and Langer, 2012]. The theory, with just a few parameters, has successfully reproduced a large number of experiments and molecular dynamics simulations for glassy materials, including strain localization patterns [e.g., Langer and Manning, 2007; Manning et al., 2007; Lieou and Langer, 2012; and the review papers by Daub and Carlson, 2010; Falk and Langer, 2011]. Recently, the theory was extended to model shear flows of granular materials with breakable particles [Lieou et al., 2014], incorporating flash heating effects [Elbanna and Carlson 2014] as well as acoustic fluidization under low normal stresses [Lieou et al., 2015a]. Furthermore, the theory has been implemented to investigate conditions for stability of sliding and strain localization in granular layers subjected to shear and vibrations [Lieou et al., 2015b; Kothari and Elbanna, 2016] and has pointed to the possible effect of vibrations in transition from rapid slip to slow slip and eventually stable sliding; a mechanism that may play a role in tremor-slow slip interaction [Lieou et al., 2016].

A basic feature of granular materials is that particles move and rearrange in response to applied stress. Molecular dynamics simulations of glassy materials [Falk and Langer, 1998] and discrete element models of granular materials [Ferdowsi et al., 2016] reveal that plastic irreversible deformation is concentrated in localized regions which came to be called shear transformation zones (STZs). These regions undergo configurational rearrangement by flipping between two orientations, anti-aligned and aligned with the direction of principal shear stresses [Falk and Langer, 1998].

A single STZ transition is represented by an irreversible rearrangement of a cluster of particles, whereby the particles locally exchange nearest neighbor relationships. While an arbitrary motion in a granular assembly may be decomposed into affine (i.e. a linear map) and non-affine components, it is the irreversible non-affine rearrangements (i.e. topological changes) of the particles that characterize an STZ. A schematic diagram of this process is shown in **Figure 1**.

The creation, annihilation, and transition of STZs may be due to mechanical forcing, acoustic vibrations, or thermal fluctuations (for nanoparticles). In this paper, we focus on athermal STZ dynamics that is mainly caused by shearing.

A state variable, the effective temperature (or compactivity) $\chi$, is introduced to measure the degree of configurational disorder in the system and to set the number of STZs. The effective temperature ( or compactivity) is formally defined as the change in the system potential energy (or volume $V$) per unit change in the system configurational entropy $S_c$ : $\chi = \partial V / \partial S_c$ [Bouchbinder and Langer, 2009; Lieou and Langer, 2012]. A fundamental result in the STZ theory is that the continuous creation and annihilation of the STZs drive their density $\Lambda$ toward a Boltzmann probability distribution $\Lambda \sim \exp(-1/\chi)$ [Langer, 2008; Langer and Manning, 2008; Bouchbinder and Langer, 2009].

The STZ transition generates an epoch of local plastic strain. The plastic strain may be decomposed into a deviatoric and a volumetric part. The local deviatoric plastic strain may vary from one STZ to another but it is taken on average to be of magnitude ($\varepsilon$). The macroscopic plastic strain is the cumulative result of many local events. The mathematical formulation of the STZ theory as a viscoplastic framework may then be summarized as follows:

**Flow rule**: The deviatoric part of the plastic strain rate is given by:

$$\mathbf{D}^p_{dev} = \tau^{-1} \Lambda^{eq} R(\bar{s}, \chi) \mathbf{s} / \bar{s} \tag{1}$$

Here, **s** is the deviatoric stress, $\Lambda^{eq} = \exp(-1/\chi)$ is the STZ density, $R(\bar{s}, \chi)$ is the rate switching function which may be interpreted as a success probability for an STZ transition and is dependent on the second invariant of the deviatoric stress $\bar{s}$ and the compactivity $\chi$, and $\tau$ is a characteristic time scale. The deviatoric strain rate tensor is taken to be co-axial with the deviatoric stress tensor. This is an adequate assumption for isotropic amorphous materials [Pechenik, 2005]

The characteristic time scale, also known as the inertial time scale, depends on pressure $P$, grain size $a$, and particle density $\rho$ and is given by: $\tau = a\sqrt{\rho/P}$ [da Cruz et al., 2005; Jop et al., 2006; Lieou and Langer, 2012]. The second invariant of the deviatoric stress tensor is given by the classical formula $\bar{s} = \sqrt{\frac{3}{2} s_{ij} s_{ij}}$. The rate switching function $R$ may be assumed based on physical arguments, derived directly from discrete element models or constrained through the second law of thermodynamics. We adopt the latter approach and follow earlier work [Bouchbinder and Langer, 2009; Lieou and Langer, 2012; Lieou et al., 2014] which showed that a thermodynamically consistent rate factor is given by:

$$R(\bar{s}, \chi) = \begin{cases} \exp(\bar{s}/P\chi)[1 - s_o/\bar{s}] & \text{if } \bar{s} > s_o \\ 0 & \text{if } \bar{s} < s_o \end{cases} \qquad (2)$$

Where $s_o$ is the minimum flow stress. The above formula indicates that no plasticity occurs if $\bar{s} < s_o$. Thus, it is possible to interpret $\bar{s} = s_o$ as the initial (local) yield surface equation. The value of $s_o$ is a function of many system variables including the grain shape, surface roughness, contact temperature, etc. Here we assume $s_o$ to linearly depend on the first invariant of the stress tensor $I_1$. That is, we define $s_o = \alpha_1 - \alpha_1 I_1/3$, in which $\alpha_1$ and $\alpha_1$ are constants.

The volumetric strain rate is directly proportional to the rate of compactivity. To see this, we start with the definition of effective temperature $\chi = \partial V / \partial S_c$ from which it follows that $V = V(S_c)$. Taking time derivatives of both sides and acknowledging that $S_c = S_c(\chi)$ (since $S_c$ must depend on the number of STZs)

$$\dot{V} = \frac{\partial V}{\partial S_c} \frac{\partial S_c}{\partial \chi} \dot{\chi} = \chi \frac{\partial S_c}{\partial \chi} \dot{\chi} = \alpha_{eff} \dot{\chi} \tag{3}$$

where $\alpha_{eff}$ is the effective volume expansion coefficient. Assuming volumetric deformation to be isotropic, the inelastic volume strain rate tensor is given by:

$$\mathbf{D}^p_{vol} = \alpha_{eff} \dot{\chi} \mathbf{I} \tag{4}$$

Here, $\mathbf{I}$ is the identity tensor. The total inelastic strain rate is:

$$\mathbf{D}^p = \mathbf{D}^p_{dev} + \mathbf{D}^p_{vol} \tag{5}$$

**Internal variable evolution:**

The evolution of compactivity $\chi$ is given by:

$$\dot{\chi} = \left(1 - \frac{\chi}{\hat{\chi}}\right) \frac{\mathbf{s} : \mathbf{D}^p}{Pc_o} + \nabla \cdot \left(D_o D^p_{eq} \nabla \chi\right) \tag{6}$$

The above equation states that only a fraction of the external work rate $\mathbf{s} : \mathbf{D}^p = s_{ij} D^p_{ij}$ is dissipated to increasing $\chi$ as it is driven toward its steady state value $\hat{\chi}$ [Langer and Manning, 2008]. This fraction is given by $1 - \chi/\hat{\chi}$. The coefficient $c_o$ sets a scale for the amount of work required to increase the compactivity. The larger $c_o$ the slower the evolution of $\chi$ is. The second term on the right-hand side of equation (6) is effective only if $\chi$ is spatially heterogeneous. It represents a diffusive process, analogous to diffusion effects in the heat flow problems, through which disorder in a certain domain is being transmitted to nearby regions. The diffusion term spatial scale $D_o$ is proportional to the square of the STZ size (~ few particles) and the diffusion time scale is set by the equivalent plastic strain rate $D_{eq}^p = \sqrt{\frac{3}{2} D_{ij} D_{ij}}$. The steady state value of the effective temperature is rate dependent [Langer, 2007; Manning and Langer, 2008; Daub and Carlson, 2009; Lieou et al., 2015]. However, at strain rates well below a critical value, given by the inverse of the inertial time scale $\tau^{-1}$, this rate dependence is weak and the steady state value may be taken as a constant $\chi_\infty$ [Lieou et al., 2014; Kothari and Elbanna, 2016]. For the parameters used in this paper, the critical strain rate is of the order of $10^7 / s$ which is orders of magnitudes higher than the strain rate of interest here. Thus, we will adopt the approximation that $\hat{\chi} = \chi_\infty$.

**Closure of the system of equations:** The flow rule and the internal variable evolution must be supplemented by a stress update rule to complete the system description and advance the solution in time. The stress update scheme depends on how we choose to decompose the deformation into elastic and inelastic components.

Here, we adopt a multiplicative decomposition [Bilby et al., 1957; Kroner, 1958; Lee, 1969] for the total deformation gradient $\mathbf{F}$ into an elastic $\mathbf{F}^e$ and viscoplastic $\mathbf{F}^p$ components: $\mathbf{F} = \mathbf{F}^e \mathbf{F}^p$ as shown in **Figure 2**.

The Cauchy stress $\boldsymbol{\sigma}$ is then given by

$$\boldsymbol{\sigma} = J^{e-1}\mathbf{F}^e : \mathbf{T} : \mathbf{F}^{e^T} \qquad (7)$$

Where $J^e$ is the determinant of the elastic deformation gradient $\mathbf{F}^e$ and $\mathbf{T}$ is the 2$^{nd}$ Piola-Kirchhoff stress computed in the intermediate configuration and is given by: $\mathbf{T} = \mathbf{L} : \mathbf{E}^e$ such that $\mathbf{L}$ is the elasticity tensor and $\mathbf{E}^e$ is the Green-Lagrangian finite strain tensor. The latter is obtained from the Cauchy Green strain tensor $\mathbf{C}^e = \mathbf{F}^{e^T}\mathbf{F}^e$ through the relation $\mathbf{E}^e = \frac{1}{2}(\mathbf{C}^e - \mathbf{I})$.

The updated value of the elastic component of the deformation gradient $\mathbf{F}^e_{n+1}$ is obtained from the multiplicative decomposition rule: $\mathbf{F}^e_{n+1} = \mathbf{F}_{n+1}\mathbf{F}^{p^{-1}}_{n+1}$, where the total deformation gradient $\mathbf{F}_{n+1}$ is based on the displacement field and $\mathbf{F}^p_{n+1}$ is computed from the viscoplastic velocity gradient: $\mathbf{L}^p = \dot{\mathbf{F}}^p\mathbf{F}^{p^{-1}} = \mathbf{D}^p = \dot{\lambda}\mathbf{r}$. Here $\dot{\lambda}$ is the plastic flow rate and $\mathbf{r}$ is the plastic flow direction, both of which are determined from the flow rule equation (1) and (4), and the plastic flow rate is updated, with the stress, using Newton-Raphson iterations. We provide a brief description of the finite deformation framework and the update algorithm in Appendix A.

## 3. Numerical implementation

We implement the STZ constitutive model within a finite deformation finite element framework provided by MOOSE platform from Idaho National Lab. MOOSE: The Multi-Physics Object Oriented Simulation Environment (Moose) [Gaston et al., 2009] provides a flexible platform to solve multi-physics problems implicitly and in a tightly coupled manner on unstructured meshes. Moose framework is built on top of libraries including the libmesh finite element library [Kirk et al., 2006] and PETSc solver library [Balay et al., 2010]. The solution of the nonlinear equations of dynamic equilibrium and material model update is done using Jacobian-free Newton-Krylov (JFNK) approach [Knoll and Keyes, 2004].

**Model Setup:** We consider a layer of granular materials sheared between two parallel planes and subjected to a constant pressure at the top and the bottom as shown in **Figure 3**. To mimic an infinite long strip, periodic boundary conditions are enforced at the left and right boundaries. At the center of the fault gouge, we introduce a perturbation in the local disorder by defining a circular inclusion with higher initial compactivity (aka less dense) than the surrounding bulk. To ensure the quality of the solution, an h-refinement study has been conducted to verify convergence with increasing resolution. The material properties that are used in the simulation are summarized in **Table 1.**

The loading in each simulation is a two-stage process. In the first loading step we solve a static equilibrium problem for the applied pressure at the top and bottom (assuming periodic boundary conditions at the sides). Then we apply the shear loading. To achieve a constant strain rate, we adopt a ramp loading technique in which we change the strain rate from zero to the prescribed constant value over a finite period of time using a $5^{th}$ order of polynomial in time. This ensures a smooth profile for the displacement, velocity and acceleration at the top and bottom boundaries throughout the loading history. Since our simulation is fully dynamic, this ramping reduces the effect of waves that may be generated due to an abrupt change in the loading rate during the elastic regime.

## 4. Results

### 4.1 Generic Shear band localization

With the circular inclusion having a slightly higher compactivity than the rest of the granular layer, plasticity starts in that zone first. Due to the interdependence between the compactivity and inelastic strain rate (equation (1)-(6)), this positive feedback causes localization of the plastic deformation in the inclusion and favors the regions in its immediate vicinity to accommodate further inelasticity. This leads to the nucleation of shear bands from the circular inclusion and their subsequent growth. As shown in **Figure 4**, the resulting localization pattern, as measured by the distribution of the compactivity, agrees well with the schematic describing field observation as summarized in Logan [2007]. In particular, our numerical results capture the Riedel "R1", X, and Y-bands.

To further explore the relation between the stress slip response and shear bands propagation, we show in **Figure 5** the evolution of the localization patterns and its correlation with different stages in the stress slip plot. Initially, the response of the layer is elastic. With the initiation of plasticity in the central inclusion, the global response does not change much. However, with progressive shearing, the region with inelastic strain grows and diagonal bands start to propagate. With the localization of plastic deformation in the growing shear bands, the local inelastic strain rate increases causing the initiation of strain softening. This continues as the diagonal bands (X and R) propagate towards the top and bottom boundaries of the sample where new boundary bands start to develop. As the stress response approaches the steady state, the through-going Y-band becomes more visible. Interestingly, the Y-band seems to emanate from the Riedel shear band by curving out of it at some point during the softening stage before becoming fully developed. This picture of the stress slip response alongside the compactivity evolution and localization bands development agrees qualitatively with the experimental observations of strength and microstructure evolution discussed in [Marone, 1998].

## 4.2 Factors affecting shear localization

### i. The effect of dilatancy

One relevant feature in the STZ formulation is that the inelastic dilatancy coefficient is not prescribed but it evolves as part of the solution. To show this let's consider the theory formulation in homogeneous 1D setting. The evolution of the compactivity reduces to:

$$\dot{\chi} = \frac{s\dot{\gamma}}{c_o p}(1-\chi/\hat{\chi}) = \frac{\dot{V}}{\alpha_{eff}}$$ It follows that the dilatancy parameter is given by:

$$\beta = \frac{\dot{V}}{\dot{\gamma}} = \frac{\alpha_{eff}}{c_o}(1-\chi/\hat{\chi})\mu$$ where $\mu = s/p$. Thus the dilatancy evolves as a function of stress, pressure and disorder. In the absence of additional constraints, we hypothesize that $\alpha_{eff}$ and $c_o$ are of the same order of magnitude and we take them to be equal.

We vary $c_o$ to investigate the effect of inelastic dilation on the stress slip response and shear band evolution. The results are shown in **Figure 6**. As the value of $c_o$ increases, the compactivity increases more slowly, the peak stress increases and more localization is observed. This behavior is characteristic of dilatant media. On the other hand, if $c_o$ becomes small enough the inelastic dilatancy is negligible and occurs almost instantaneously and the subsequent response becomes more ductile like with no noticeable strain localization.

### ii. The effect of Ramping Protocol (Inertia Effect)

In this section, we investigate the possible inertia effects induced by the ramping the shear loading differently from the default case. We fix the steady state imposed strain rate value as well as the fifth order polynomial used for strain rate interpolation but we progressively increase the time period over which the ramping occurs. As shown in **Figure 7**, with the increase of the ramp loading period, the shear band is more distributed, and the Y band development is less distinct than in the default case. Furthermore, with increasing the ramping time, the elastic regime becomes smoother due to the absence of the small stress perturbations carried by the propagating waves. The results suggest that, while the inertia effect doesn't change the overall qualitative features of the strain localization pattern, the Y-band is more mature under quicker ramping. A possible explanation for this is that under quicker ramping, waves emanate from the top and bottom boundaries and interfere constructively in the middle of the layer causing stronger localization. This may have implications for co-seismic strain localization in which the strain rates increase rapidly at the rupture tip possibly favoring the Y-band formation more. The growth of the X band is not visibly affected by the ramping rate.

### iii. The effect of loading rate

Here, we use the same ramping protocol as in the default case but we change the value of the steady state imposed strain rate. **Figure 8** shows the distribution of effective temperature under three different strain rates. As loading rate increases, the boundary shear bands widen. Furthermore, the bifurcation of the Riedel band and Y bands is developed earlier. Therefore, comparing the three loading rates at fixed time, suggests that the plastic strain accumulated by the through-going bands (boundary and Y bands) is higher at higher strain rates.

### iv. The effect of confining pressure

In this section, we vary the applied confining pressure $P$ on the top and bottom of the fault gouge specimen, and study its effect on strength and shear and evolution. **Figure 9** shows that with the increasing confining pressure, the specimen exhibits a brittle to ductile transition. We define a brittle behavior by the existence of a strength drop and localized deformation, whereas in ductile response these features are absent. The observed transition as a function of pressure may be explained as follows. At higher pressures, the minimum flow stress $s_o$ increases. This delays the initiation of plasticity and increases the peak stress as well as the steady state flow stress. However, with increasing pressure, the characteristic time scale for STZ transition $\tau$ decreases (equation (1)) enabling faster accumulation of inelastic strains and disorder (similar to the effect of reducing $c_o$ discussed previously). At low pressure, the plasticity accumulates slowly causing the stress to peak followed by strain softening and brittle behavior. At high pressure, the plasticity accumulates quickly after its initiation and is distributed across the layer leading to gradual saturation of the strength without softening signatures. This is also reflected in the shear band plots where strain localization is evident at low pressures but shear bands are diffusive and distributed across the sample width at higher pressure.

**v. The effect of initial compactivity**

The initial preparation of the sample may affect its subsequent response. In another amorphous system, namely bulk metallic glasses, it was shown that a well-aged sample exhibits a brittle response whereas a more disordered sample exhibits a ductile response [Rycroft and Bouchbinder, 2012]. A similar observation has been documented for granular materials where the initial relative density plays a similar role to the degree of ageing. That is, an initially dense sample exhibits a brittle response whereas an initially loose sample is ductile [Marone, 1998; Silvio et al. 2008; Andre et al. 2009, Samual, et al. 2012; Carpenter et al. 2015]

Here, we examine the effect of initial disorder on the brittle to ductile transition in sheared confined layers. We consider two cases: one with the default background compactivity $\chi = 0.04$ and the other with higher initial background compactivity $\chi = 0.06$ (closer to steady state value of 0.08). The results are shown in **Figure 10**, where we plot the stress strain response as well as snapshots of the final compactivity distribution. For higher initial compactivity, and despite the existence of the central inclusion, it is more favorable for the sample to distribute plastic strain across the whole layer. Even if the inelastic deformation starts from the central inclusion, the disorder is high enough everywhere to accommodate plasticity shortly after. This leads to a ductile behavior in the sense that the stress progressively increases towards steady state without exhibiting a peak or strain softening. Furthermore, the plastic deformation is well distributed across the layer with no visible localization. On the other hand, with lower initial compactivity, the behavior is brittle. The inelastic deformation is localized in the center inclusion and the shear bands that grow and propagate out of it. The stress peaks at a certain slip and then goes through strain softening phase before eventually reaching a steady state value.

Compactivity $\chi$ is related to porosity [Lieou et al., 2014] and changes in compactivity relates to changes in pore volume as discussed earlier (assuming incompressibility of the granular particles). Higher initial compactivity corresponds to higher initial porosity. Thus, our results suggest that initially loose layers exhibit brittle response while initially dense layer exhibit a ductile one. The brittle to ductile transition as a function of initial preparation agrees with some experimental observations [Marone and Scholz, 1989].

We have also tested the response corresponding to a random distribution of the initial compactivity. We set the initial compactivity as a random variable that is uniformly distributed between 0.04 and 0.05. The results are shown in **Figure 11**. The localization pattern is richer than the default case and exhibit more complexity. Conjugate shear bands emanate from multiple nucleation sites and interact with each other resulting in a complex localization texture. This pattern is qualitatively similar to the localization bands observed in experiments on analogue materials as well as in natural fault zones [ Mair and Marone, 1999; Logan, 2007; Haines, et al. 2012].

**vi. The effect of layer thickness**

It is natural to expect that the localization pattern depends on the layer thickness since the shear band has a finite length scale and for narrow enough layers, boundary interactions become important. We investigate the evolution of shear bands for three values of the layer thickness: 0.1m, 0.2m, and 0.3m, sheared at the same loading velocity. As shown in **Figure 12**, for the smallest thickness, the inelastic deformation almost fully saturates the layer and there is no distinct Riedel band. At intermediate thickness, there is more space for the Riedel band to grow and reach the boundary. The Y-band also starts to develop. Boundary interactions still exist as evident by the development and growth of boundary shears at the right bottom and top left edges. For the largest thickness considered, the Y-band bifurcates at an earlier stage. The Riedel shear continues to grow towards the boundaries. However, the development of the boundary shears is delayed due to the absence of interaction of the boundaries with either the central inclusion or the Riedel bands.

## 5. Discussion

Understanding deformation and failure in granular materials is a problem of both fundamental importance and practical relevance. This is because many natural phenomena as well as industrial processes are controlled by the physics of granular deformation. Earthquakes, and landslides, as well as pouring, transportation, and mixing in food and pharmaceutical industries are just few examples where granular rheology, especially that which involves the shear response due to local particle rearrangement, is of direct relevance. In many circumstances, the granular deformation is not macroscopically uniform but localizes in shear bands. This is particularly the case for crustal faults in which amble evidence exist that co-seismic deformation localizes in thin regions within broad damage zones [Rice, 2006]. In this paper, we presented a numerical model for the evolution of viscoplastic deformation in a sheared granular layer under constant pressure with spatially heterogeneous porosity-like parameter. Using the model, it was possible to investigate the strength evolution and shear band development under different loading conditions.

The viscoplastic formulation adopted here is based on the shear transformation zone (STZ) theory, a non-equilibrium statistical thermodynamic framework for describing rate dependent inelastic deformation in amorphous materials. The STZ theory belongs to the broader class of constitutive laws with internal state variables [Rice, 1971]. Only the initial yield surface is defined but subsequent hardening or softening is computed as part of the solution by integrating the evolution equations of the internal variables. There have been prior fundamental work for investigating localization and plasticity in pressure sensitive materials [See for example, Rudnicki and Rice, 1975; Rice and Rudnicki, 1980; Borja, 2004; Borja and Andrade, 2006; Andrade and Borja, 2006; Poulet and Veveakis, 2015]. Most of the prior work has implemented phenomenological constitutive models. One point of departure in the current work is that the primary internal variable, the compactivity, has a clear interpretation as a measure of local disorder in the system and possesses a formal connection to fundamental thermodynamic quantities such as volume and entropy. Furthermore, the flow rule in the STZ formulation is consistent with a microscopic picture of transition dynamics of defects (STZs flipping and sliding). These features make the STZ distinct from other widely used plastic models such as Cam-Clay, Mohr-Coulomb and Drucker-Prager models and enable direct connection with small scale molecular dynamics or discrete element models.

A widely used constitutive model for friction in rocks and granular materials is the rate and state law [Dieterich 1979; Ruina, 1983]. Despite its phenomenological nature, the rate and state formulation has led to significant progress in describing several sliding phenomena particularly in bare rock surfaces experiments. The inclusion of the slip rate history, through state variables, in addition to the instantaneous slip rate in evaluating the friction coefficient has been a leap forward from earlier friction laws (see, for example, the discussion in [Rice et al., 2001]). However, the lack of physical interpretation of the state variables in the Dieterich-Ruina laws limits their predictive capability especially when it comes to gouge layers. While in principle, it is possible to include more than one state variable to capture increasingly complex behavior, constraining the evolution equations of these internal variables is not straightforward. An advantage of the STZ theory in that context is that it is based on the laws of thermodynamics and the state variable evolution is constrained by energy flow and entropy evolution. In the past few years, the STZ framework has been successfully extended to incorporate physical phenomena critical for gouge mechanics such as grain fragmentation [Lieou et al., 2014], flash heating [Elbanna and Carlson, 2014], dilation and compaction under combined shear and vibration [Lieou et al., 2015, 2016]. Furthermore, extending STZ theory to higher dimensions (2D and 3D) is possible, enabling investigation of strain localization and inhomogeneous plastic deformation. It is not obvious how the rate and state friction may be consistently generalized to higher dimensions, despite some prior notable attempts [e.g. Sleep et al., 2000]

While the STZ theory provides a powerful framework for describing the multiphysics of spatially extended gouge layers, the physics-based nature of the theory comes at a price: there are more parameters to be constrained in the STZ theory than in the rate and state framework. Fortunately, many of these parameters may be constrained based on physical arguments or from classical experiments. For example, Daub and Carlson [2009] have shown, using a simple block slider setup, that it is possible to derive quantities analogous to the direct effect, slip evolution distance, and rate sensitivity parameter, by taking various partial derivatives of the STZ equations at both the transient and steady state limits with respect to strain rate. Using this procedure, experiments like velocity stepping as well as slide-hold-slide tests may be used to constraint several STZ parameters. Other parameters may be freely adjusted to fit experimental observations [see Table 1] or derived directly from subscale discrete element models.

The theoretical basis of the STZ framework makes it also possible to connect to more detailed microscopic models such as molecular dynamics and discrete element models. Recently, molecular dynamics simulations of glassy systems have made significant progress in identifying STZs using analysis of soft modes [Manning and Liu, 2011] or susceptibility of molecular clusters to yielding [Patinet, et al. 2016]. By counting the number of STZs in a given volume, the effective temperature may be directly computed using the Boltzmann distribution in STZ density. Thus, a quantitative measure of the effective temperature may be established. Furthermore, microscopic models may help in constraining parameters in the rate factors as well as those connecting the compactivity and volume changes (dilatancy). In principle, it may be possible to do more detailed studies using subscale discrete element models that communicate with the quadrature points of our finite elements. This integrated approach will provide a truly multiscale and predictive formulation for gouge deformation and is a candidate for future work.

The STZ formulation does not just compare well to rate and state friction in capturing basic features of sliding response [Daub and Carlson, 2009] but it may also highlight additional physics that may help modify the rate and state laws. For example, the internal state variable in the STZ formulation, the compactivity, is driven by the inelastic work rate and not only by the slip (or strain) rate as in the classical evolution laws for the rate and state friction. The dependence of the compactivity on the work rate follows directly form the first law of thermodynamics and suggests that the evolution of the state variable should depend on the stress as well as the strain rate. This remains to be tested in friction experiments by carrying stress stepping tests in addition to velocity stepping ones. Furthermore, there has been some previous notable attempts for incorporating porosity evolution in the classical rate and state law and accounting for inelastic dilatancy [Sleep 1995; Segall and Rice, 1995]. In this prior work the dilatancy factor was assumed to be constant. In the current work, the dilatancy evolves as part of the solution and depends on pressure, stress and disorder.

In this paper, we have investigated stress slip response and shear band evolution in sheared granular layers under different conditions of confining pressure, dilation, and loading rates. We have shown that our numerical predictions agree qualitatively with many generic features of gouge deformation reported lab and field observations such as different shear band orientations, brittle to ductile transition with increasing confining pressure, brittle to ductile transition as a function of initial porosity, and increase in the peak strength with increased dilatancy. However, the range of parameters explored here was rather limited. For example, the confining pressure was only increased from 10 MPa to 20 MPa. Strain rates were also varied only between 4/s to 16/s. Seismogenic conditions may require testing gouge response up to hundreds of MPa of confining pressure and at strain rates up to 100 or 1000/s. The range of parameters considered in this study is thus closer to experimental conditions than to field conditions. Extension to high pressures and strain rates will be the focus of future investigations.

In the last three decades, significant progress has been achieved in testing gouge layers at different pressures and loading rates. However, challenges still exist when it comes to mimicking conditions prevailing during earthquakes. For example, the increase in temperature at large slip velocities and pressures may melt the machine rim. Also, it is hard to confine the gouge layer at high slip rates. Furthermore, most high speed frictional experiments are of the rotary type [Shimamoto and Tsutsumi, 1994; Di Toro et al., 2004; Han et al., 2007; Golsby et al., 2008; Di Toro et al., 2011;]. This prevents capturing physics associated with rupture propagation or inhomogeneity of slip conditions. The model explored in this study will contribute to closing this knowledge gap. By extending prior work, that has been done in the context of 1D shear zone, through including additional physics related to grain fragmentation [Lieou et al., 2014], flash heating [Elbanna and Carlson, 2014], and pore fluid pressurization [Rice et al., 2015] to the current 2D viscoplastic formulation it will be possible to predict gouge response under extreme condition, capture extreme localization and investigate competition between gouge dilatancy (captured in the current model) and pore fluid thermal pressurization.

In this paper, we have considered dry granular layers as a first step. Given that most fault zones are fluid infiltrated, it is important to couple the current viscoplastic formulation with an equation for pore fluid pressure evolution in response to gouge volume changes. Furthermore, at high strain rates, shear heating will be high enough to cause constrained expansion of pore fluids leading to thermal pressurization. Future work will focus on integrating temperature and pore pressure evolution in the current model. Moreover, our recent work [Lieou et al, 2015, 2016; Kothari and Elbanna, 2016], inspired by experiments of Van der Elst et al., 2012 and discrete element models by Ferdowsi et al., 2015, suggest that acoustic vibrations may cause transient compaction, alter the stability of sliding in frictional fault gouge, trigger slip, and cause strain delocalization. Extending these results to higher dimensions (2D and 3D) will allow exploring the effect of vibrations on slip, and localization in spatially heterogeneous conditions which may have important implications for triggered earthquakes and slow slip [Lieou et al, 2016]. Finally, we acknowledge that fault zones have 3D structures. The current plane strain formulation can be extended to 3D in a straightforward way. However, the computational cost are orders of magnitude higher. We will report on our ongoing efforts in running a 3D implementation of our numerical method on the National Petascale Computing Facility Blue Waters [Bode, et al., 2012; William et al., 2015] elsewhere.

## 6. Conclusion

In this paper, we present a numerical model for shear defamation in gouge based on finite deformation kinematics and the Shear Transformation Zone (STZ) viscoplasticity framework. Our numerical model generically predicts complex shear band localization pattern similar to what is reported in lab and field observations [ Scott et al., 1994; Logan et al., 1979, 2007]. Our conclusions are summarized as follows:

1- Complex strain localization patterns emerge, with minimum assumptions, including Riedel, boundary, and Y-bands. The Riedel band follows the direction of the optimally oriented shear plane and emanate first from the initial disorder perturbation. With increase slip, the Y-band starts to emerge and propagate. The boundary shear bands emerge due to the interaction of the Riedel shear and the sample boundaries.

2- With increased loading rate, more inelastic deformation is localized in the through-going shear bands including the boundary and Y-bands.

3- With increased dilatancy, the peak strength increases and the response becomes more brittle. At negligible dilatancy, the response is ductile and no shear bands form.

4- With increased pressure, both the peak strength and the steady state flow stress increase. However, the response shows a brittle to ductile transition with increased pressure.

5- With increased initial disorder (aka increased initial porosity) a brittle to ductile transition is also observed. For initially loose layer, the shear stress progressively increases towards the steady state value and the plastic deformation is distributed across the layer. For initially dense layers, the shear stress reaches a peak followed by strain softening and a complex shear band pattern emerges.

6- The full development of the Riedel, boundary, and Y-shears require a thick enough gouge layer. In a thin layer (where the thickness is of roughly the same magnitude as the initial perturbation) the plastic deformation is distributed across the layer and the response is ductile. With progressively increasing the layer thickness, at a constant imposed slip rate, the Riedel and Y-bands develop.

# Appendix A. Finite deformation framework

In this appendix, we briefly review the finite strain framework.

We have adopted a multiplicative decomposition by introducing an intermediate configuration (As shown in **Figure 2**), such that:

$$\mathbf{F} = \mathbf{F}^e \mathbf{F}^p \tag{A1}$$

With Jacobians $J = |\mathbf{F}|$, $J^e = |\mathbf{F}^e|$ and $J^p = |\mathbf{F}^p|$. Accordingly $J = J^e J^p$.

The gradient of the velocity field:

$$\mathbf{L} = \dot{\mathbf{F}} \mathbf{F}^{-1} \tag{A2}$$

From equation (A1) and (A2), we can express velocity gradient $\mathbf{L}$ as:

$$\begin{aligned}\mathbf{L} &= \left(\dot{\mathbf{F}}^e \mathbf{F}^p + \mathbf{F}^e \dot{\mathbf{F}}^p\right) \mathbf{F}^{p\text{-}1} \mathbf{F}^{e\text{-}1} = \dot{\mathbf{F}}^e \mathbf{F}^{e\text{-}1} + \mathbf{F}^e \dot{\mathbf{F}}^p \mathbf{F}^{p\text{-}1} \mathbf{F}^{e\text{-}1} \\ &= \mathbf{L}^e + \mathbf{F}^e \mathbf{L}^p \mathbf{F}^{e\text{-}1}\end{aligned} \tag{A3}$$

We may write $\mathbf{L}^e = \mathbf{D}^e + \mathbf{W}^e$ and $\mathbf{L}^p = \mathbf{D}^p + \mathbf{W}^p$, where $\mathbf{D}$ is the symmetric component and $\mathbf{W}$ is the skew symmetric component. We assume that $\mathbf{W}^p = 0$ [Gurtin et al., 2013]. The velocity gradient thus takes the form:

$$\mathbf{L} = \mathbf{L}^e + \mathbf{F}^e \mathbf{D}^p \mathbf{F}^{e\text{-}1} \tag{A4}$$

We further decompose the plastic strain rate tensor $\mathbf{D}^p$ into two components: a deviatoric component $\mathbf{D}^p_{dev}$ and a volumetric component $\mathbf{D}^p_{vol}$ such as:

$$\mathbf{D}^p = \mathbf{D}^p_{dev} + \mathbf{D}^p_{vol} \tag{A5}$$

Each component of plastic strain rate tensor is expressed as a product of a scalar plastic strain rate value and the plastic flow direction.

**The deviatoric component:**

$$\mathbf{D}^p_{dev} = \dot{\lambda}_{dev} \mathbf{r}_{dev} \tag{A6}$$

Where for the deviatoric plastic strain rate $\dot{\lambda}_{dev} = \tau^{-1} \Lambda^{eq} R(\bar{s}, \chi)$ and the deviatoric flow direction $\mathbf{r}_{dev} = \mathbf{s}/\bar{s}$.

**The volumetric component:**

$$\mathbf{D}_{vol}^{p} = \dot{\lambda}_{vol}\mathbf{r}_{vol} \tag{A7}$$

Where for the volumetric plastic strain rate $\dot{\lambda}_{vol} = \alpha_{eff}\dot{\chi}$ and the volumetric flow direction $\mathbf{r}_{vol} = \mathbf{I}$, the identity matrix.

We solve the plastic strain rate for each part using Newton-Raphson iterations, we form the residual $R$ for each plastic strain rate as:

$$R_{j+1}^{i} = \dot{\lambda}_{j}^{i} - \dot{\lambda}_{j+1}^{i} \tag{A8}$$

Where $i$ represent the number of plastic strain rate component (deviatoric component and volumetric component), $j$ represents the Newton-Raphson iteration counter. Then a Jacobian has to be formed by taking the derivative of the residual $R$ with respect to the plastic strain rate $\dot{\lambda}$, for simplicity, we neglect the superscript and the subscript:

$$\frac{dR}{d\dot{\lambda}} = \mathbf{I} - \frac{d\dot{\lambda}}{d\mathbf{T}}:\frac{d\mathbf{T}}{d\dot{\lambda}} \tag{A9}$$

Where $\mathbf{T}$ is then 2$^{nd}$ Piola-Kirchhoff stress. After the convergence of the plastic strain rate $\dot{\lambda}$, we obtain the updated plastic deformation gradient $\mathbf{F}_{n+1}^{p}$. Through the multiplicative decomposition of the total deformation gradient, which is calculated from the displacement field $\mathbf{F} = \frac{\partial}{\partial \mathbf{X}}(\mathbf{X}+\mathbf{u}) = \mathbf{I} + \frac{\partial \mathbf{u}}{\partial \mathbf{X}}$, we may obtain the updated elastic deformation gradient $\mathbf{F}_{n+1}^{e}$:

$$\mathbf{F}_{n+1}^{e} = \mathbf{F}_{n+1}\mathbf{F}_{n+1}^{p^{-1}} \tag{A10}$$

The elastic right Cauchy-Green deformation tensor $\mathbf{C}^{e}$ and the Green-Lagrangian finite strain tensor $\mathbf{E}^{e}$ are calculated using equations (A11) and (A12).

$$\mathbf{C}^{e} = \mathbf{F}^{e^{T}}\mathbf{F}^{e} \tag{A11}$$

$$\mathbf{E}^{e} = \frac{1}{2}\left(\mathbf{C}^{e} - \mathbf{I}\right) \tag{A12}$$

The 2$^{nd}$ Piola-Kirchhoff stress $\mathbf{T}$ is updated in the intermediate configuration through equation (A13):

$$\mathbf{T} = \mathbf{L}:\mathbf{E}^{e} \tag{A13}$$

Where $\mathbf{L}$ is the elasticity tensor. In this paper we have assumed it to be a constant and thus (A13) is just a statement of Hooke's law. A hyperelastic model can be also adopted in this formulation, if needed, to capture more complex elastic behavior. Since the 2$^{nd}$ Piola-Kirchhoff stress $\mathbf{T}$ is calculated in the unroated intermediate configuration, the objectivity is maintained.

And then we push it to the current configuration using the elastic deformation gradient and the relation between Cauchy stress $\boldsymbol{\sigma}$ and Pk2 stress $\mathbf{T}$ :

$$\boldsymbol{\sigma} = J^{e^{-1}} \mathbf{F}^e : \mathbf{T} : \mathbf{F}^{e^T} \tag{A14}$$

The algorithm for the plastic strain rate update (material model update) is summarized in the flow chart (**Figure A**)

# Acknowledgements

This research was funded by NSF/USGS Southern California Earthquake Center, funded through NSF Cooperative Agreement EAR0529922 and USGS Cooperative Agreement 07HQAG0008, and by grants from the National Science Foundation (CMMI-1435920) and (EAR-1345108).

# Tables

Table 1. Material Constants and STZ Parameters (for the default case)

| Parameter | Description | Value | Remark |
|---|---|---|---|
| $E$ | Young's Modulus | 7GPa | Range: 100 MPa (loose aggregate) - 70 GPa (grain scale). |
| $\upsilon$ | Poisson ratio | 0.4 | Range: 0.2-0.45 [Bowles, 1996] |
| $h$ | Layer thickness | 0.2m | Arbitrary |
| $w$ | Width of the fault gouge | 2.0m | Arbitrary |
| $\alpha_1$ | Cohesion | 0.0 | Neglected cohesion |
| $\alpha_2$ | Tangent of the internal friction angle | 0.6 | Typical value for sand (www.geotchdata.info) |
| $\rho$ | Material density | 1600 kg/m$^3$ | [van der Elst et al, 2012] |
| $\varepsilon$ | STZ plastic strain | 1.0 | Corresponding to a particle sliding a distance equal to its diameter |
| $c_o$ | Effective heat capacity | 0.025 | Range for dilative materials: 0.001~1.0 leading to volumetric strains between 0.01%-10% [Lieou et al., 2014] |
| $\alpha_{eff}$ | Effective volume expansion coefficient | 0.025 | Taken as same value as $c_o$ |
| $\hat{\chi}$ | Steady-state dimensionless compactivity | 0.08 | Material system specific. May be adjusted to fit experiments on steady state dilation. Maximum theoretical value is 2.57 based on the maximum bound for the STZ density [Bouchbinder and Langer, 2009; Lieou et al., 2014] |
| $\chi_o$ | Initial Compactivity | 0.04 | Depends on initial preparation. Adjustable parameter. Range: $(\ln a^3)^{-1}$ (roughly 1 STZ/m$^3$) – 2.57 (based on max STZ density) Value similar to what was used in [Lieou et al., 2014] |
| $a$ | Grain size | $1.0 \times 10^{-4}$ m | Constrained by sample particle size distribution |
| $P$ | Confining pressure | 10 MPa | Constrained by experiments or depth |

**Table 1.** List of parameters used in the simulations for the default case. Simulation with different initial compactivity $\chi_o$ confining pressure $P$ and the volume expansion coefficient $c_o$ were also carried out and are indicated in the respective sections

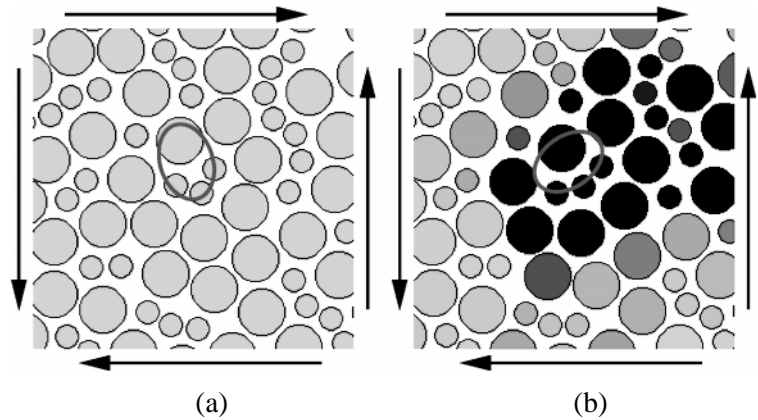

**Figure 1** Schematic of the shear transformation zone and its transition. [Reprint with permission from Falk and Langer [1998]]. (a) Oval indicates a STZ with four particles. (b) Darker particles show the reconfiguration of the STZ site after an event. The particles inside the STZ flip to another direction. This STZ will no longer be activated in the same direction but may be annihilated or flip back in the reverse direction. Other STZs will be annihilated, created and activated as well.

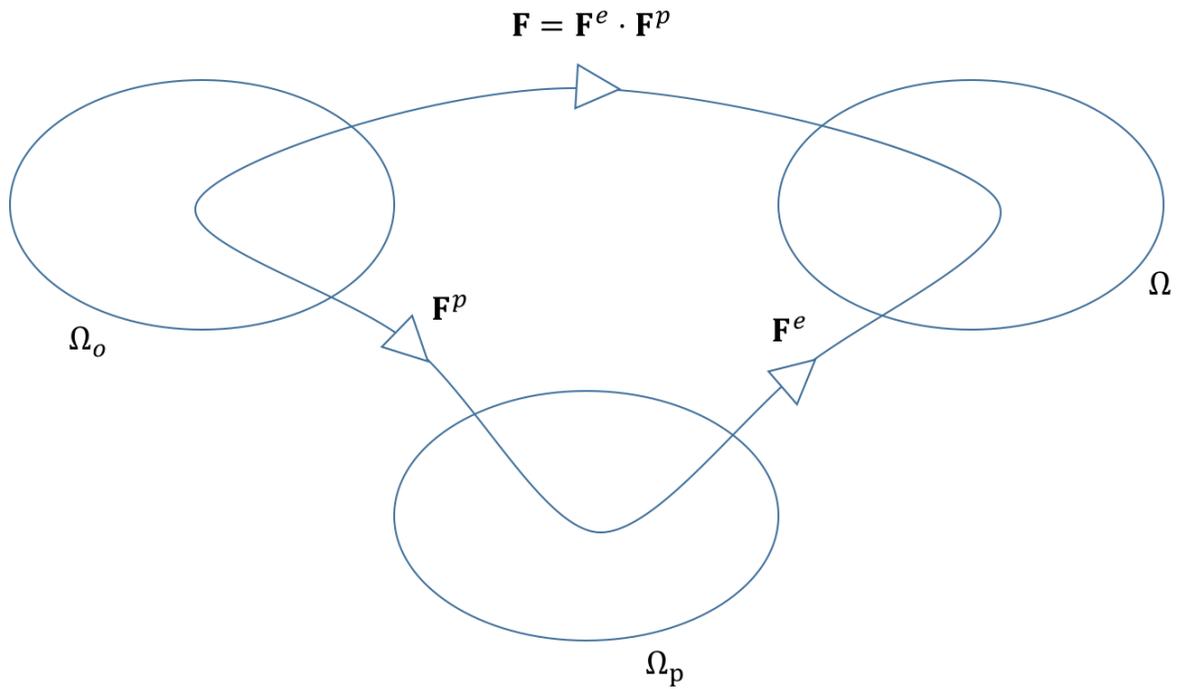

**Figure 2**. Schematic drawing of the multiplicative decomposition of the deformation gradient. The deformation gradient $\mathbf{F}$ is decomposed into an elastic part $\mathbf{F}^e$ and a plastic part $\mathbf{F}^p$ through the introduction of a fictitious intermediate configuration $\Omega_p$. Domain $\Omega_o$ represents the undeformed (reference) configuration $\mathbf{X}$ and domain $\Omega$ represents the deformed (current) configuration $\mathbf{x}$. Through the action of the deformation gradient: $\mathbf{x} = \mathbf{F}^e\mathbf{F}^p\mathbf{X} = \mathbf{F}^e\left(\mathbf{F}^p\mathbf{X}\right)$, an infinitesimal element in the reference configuration will first deform plastically in the intermediate configuration and then will rotate and stretch elastically to reside in the current configuration

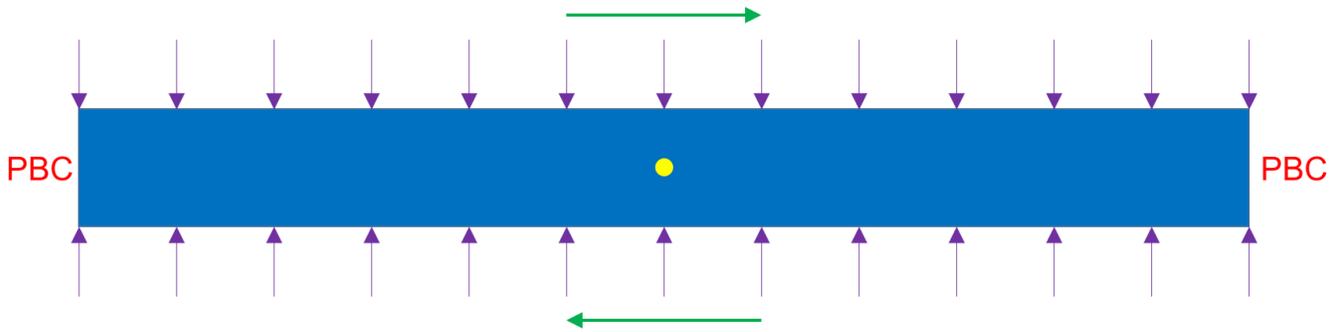

**Figure 3**.The setup of the simulated gouge layer. The layer is 2.0 m long and 0.2 m wide. Pressure is applied on the top and bottom surfaces. Shear loading is applied on the top and bottom boundaries. The shear loading is right lateral as indicated by the green arrow. Periodic boundary conditions are imposed on the lateral edges. The yellow circle is an inclusion introduced to trigger localization by having a higher initial compactivity than the background fault gouge (indicated by blue color).

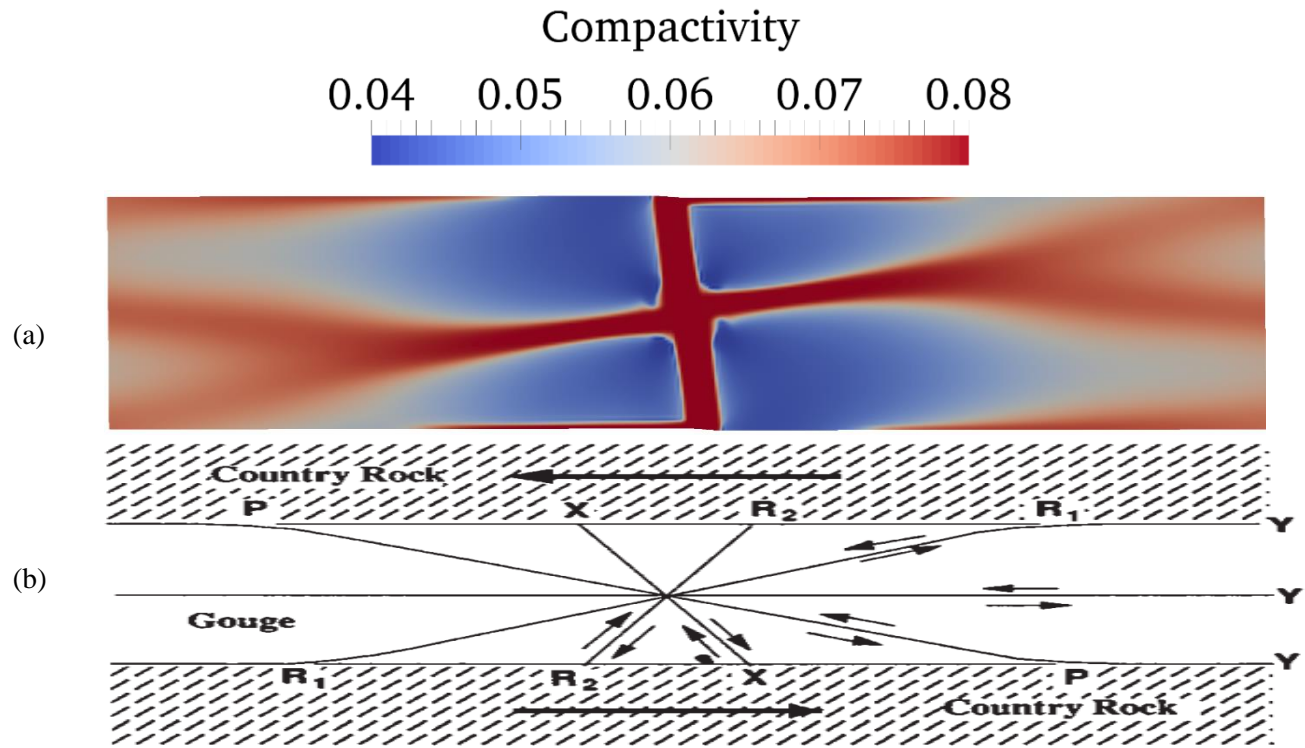

**Figure 4.** Results of strain localization pattern predicted by our numerical simulations compared to observations. (a) The distribution of compactivity at steady state from the numerical simulation. (b) Schematic drawing of experimental shear band localization pattern. [Logan, 2007]. There is a qualitative agreement between the two sets of observations. In particular, the numerical simulations generically capture the "X, R1, Y" shear bands. (Simulation Parameters: $\chi_o = 0.04$, $c_o = 0.025$, $P = 10$ MPa)

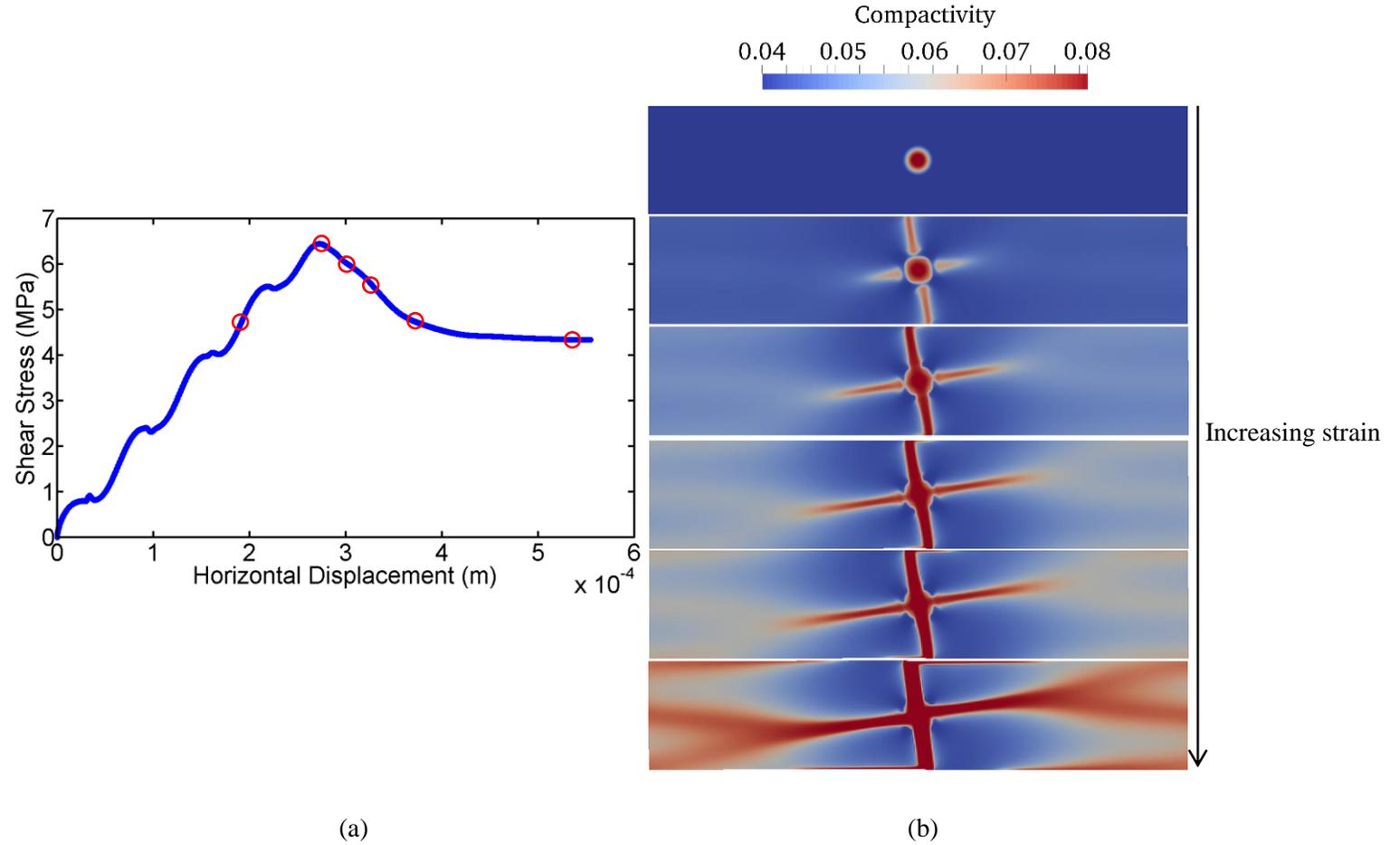

(a)              (b)

**Figure 5.** Fault gouge strength evolution and the shear band formation. (a) Stress slip response. (b) The distribution of compactivity at successive time steps corresponding to the order of the red circles on (a). At first the specimen is deforming elastically. With the initiation of the shear band from the center inclusion, diagonal bands start to form, and grow to the upper and lower boundary forming the X and Riedel shear band, and then the Riedel shear band bifurcates to Y bands which fully develop near steady state.
(Simulation Parameters: $\chi_o = 0.04$, $c_o = 0.025$, $P = 10$ MPa )

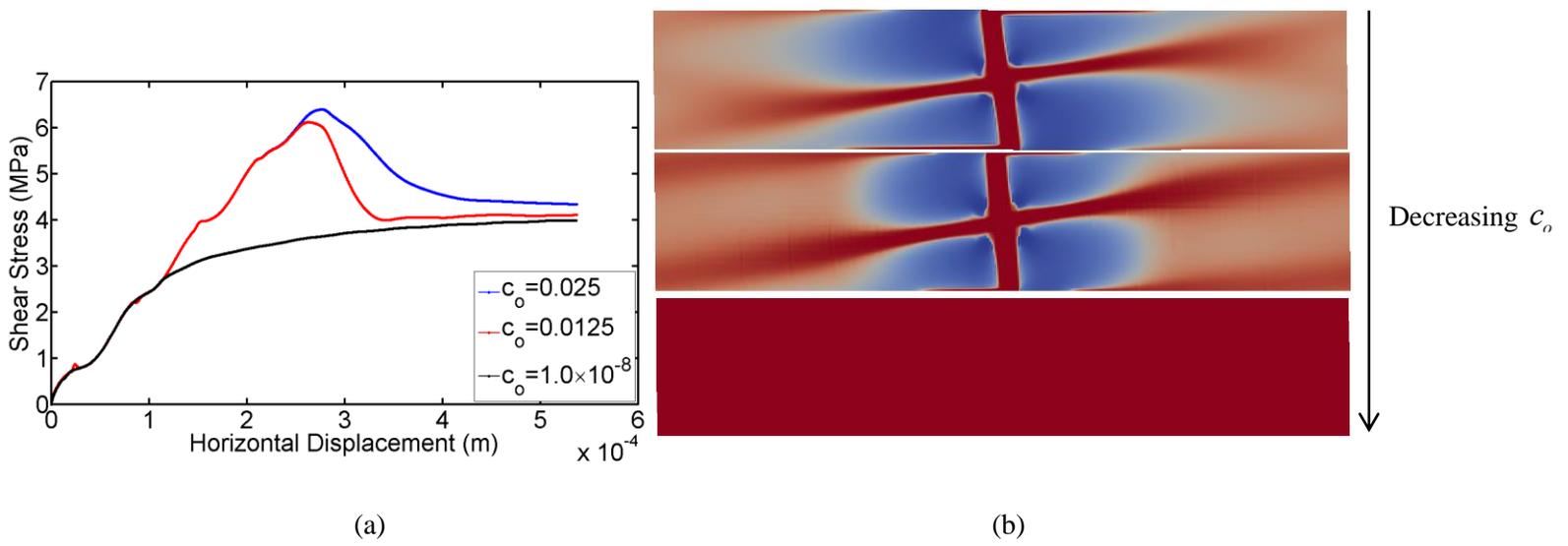

(a)                                (b)

**Figure 6**. Dilatancy effect on strength evolution and strain localization. (a) The stress slip response for different $c_o$ (b) Distribution of compactivity $\chi$ for different $c_o$ at final slip. With decreasing value of $c_o$, the specimen shows a ductile behavior with no noticeable strain localization is formed when $c_o$ is negligible. (Simulation Parameters: $\chi_o = 0.04$, $P = 10\,\text{MPa}$)

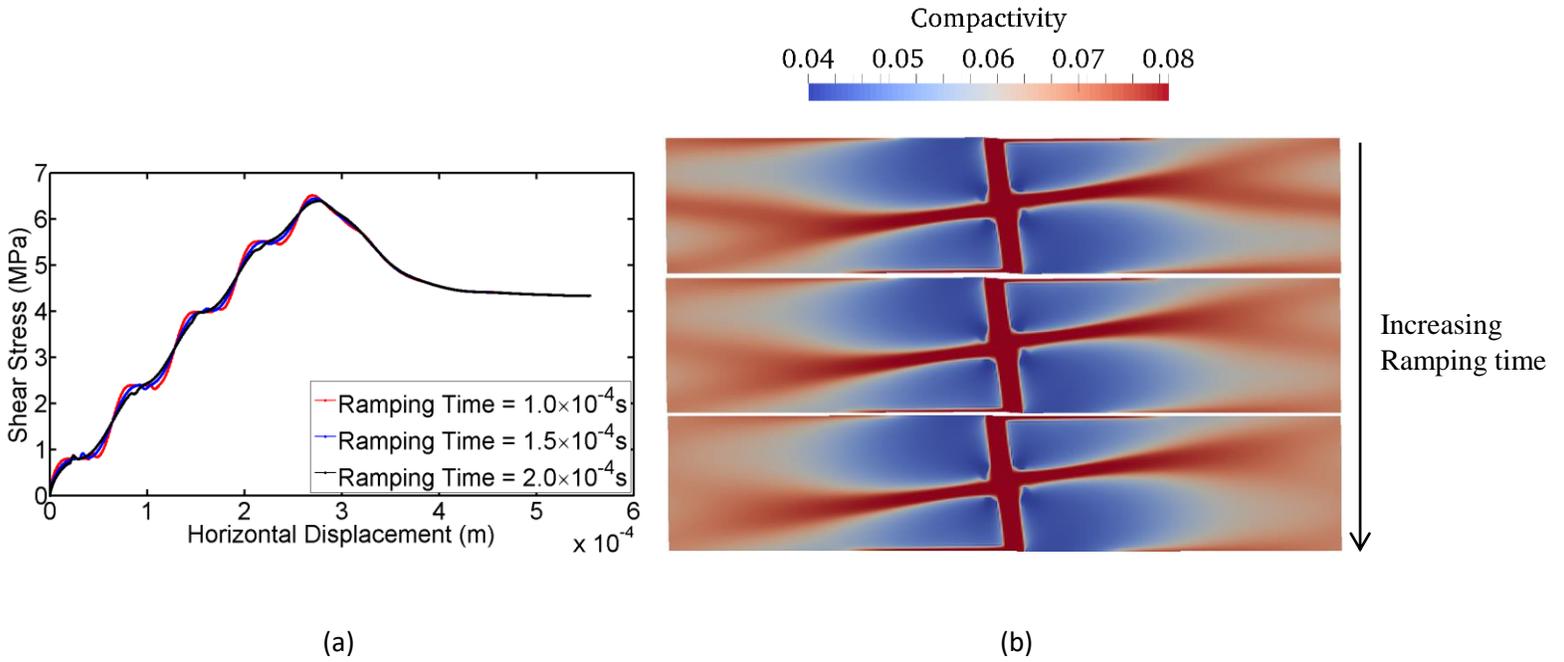

(a)                                                              (b)

**Figure 7.** Effect of ramping rate on the shear localization. (a) The shear stress slip curve with different ramping time of the imposed strain rate (steady state value = 4/s). (b) The corresponding distribution of compactivity at final slip. With shorter ramping time, the bifurcation from Riedel band to Y band is more observable. Longer ramping time causes smoother elastic response and the Y shear band is not as well developed as with the small ramping time. (Simulation parameters: $\chi_o = 0.04$, $c_o = 0.025$, $P = 10\,\text{MPa}$)

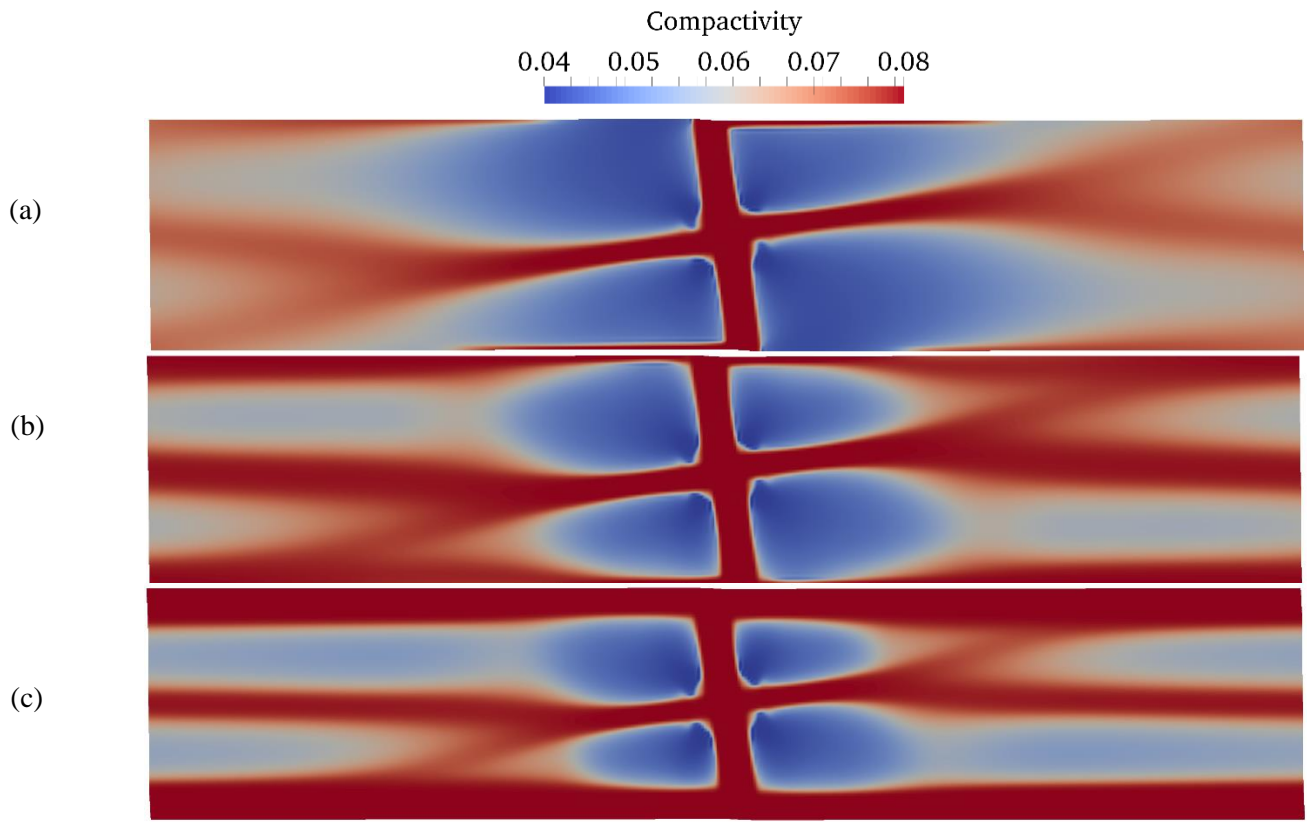

**Figure 8.** Loading rate effect on the shear band localization. Imposing Strain rate (a) 4/s ,(b) 8/s , (c) 16/s. (Simulation Parameters: $\chi_o = 0.04$, $c_o = 0.025$, $P = 10\,\text{MPa}$ )

.

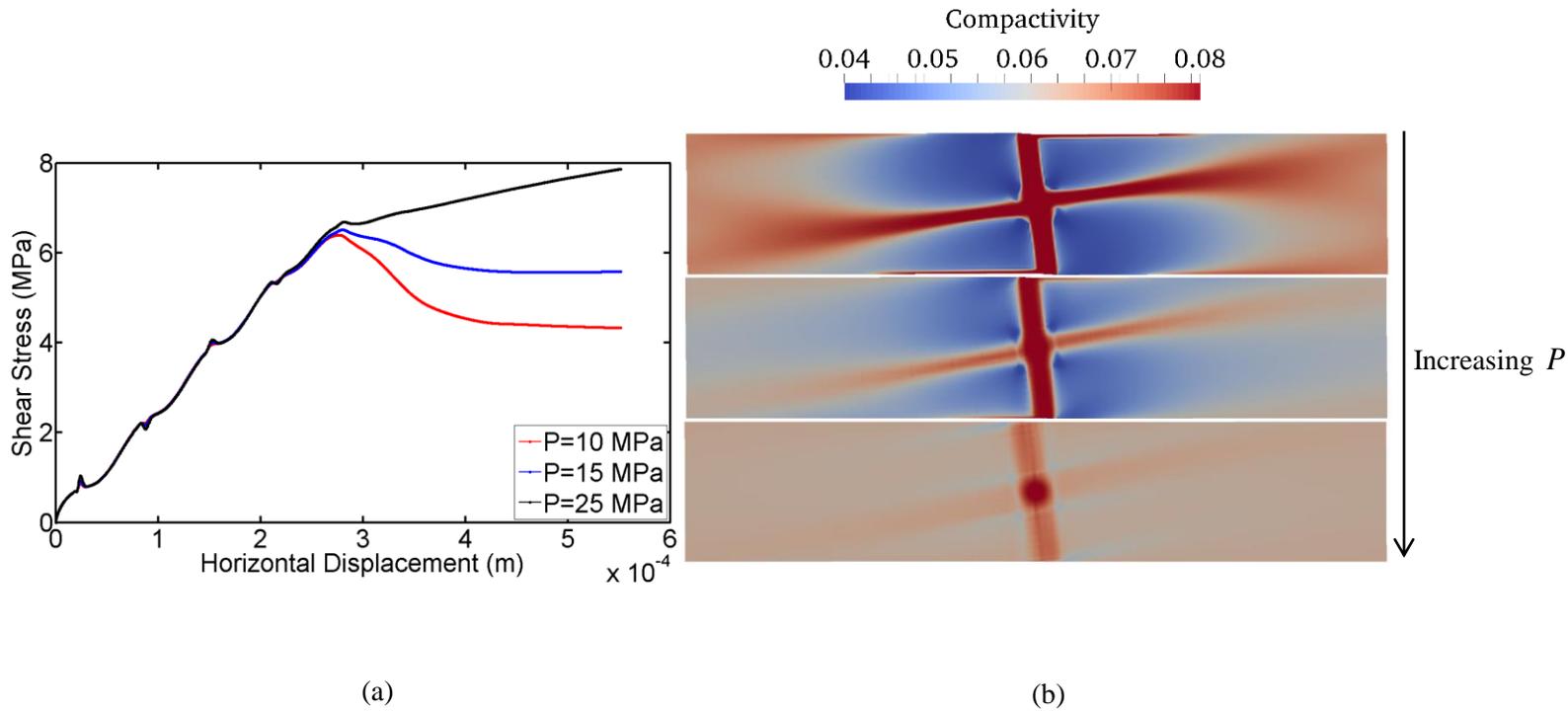

(a)                            (b)

**Figure 9.** Effect of confining pressure on shear localization. (a) Shear slip response with different confining pressure values 10MPa, 15MPa, 25MPa. With increasing pressure, the peak and flow stress increase while the strength drop decreases (b) The compactivity distribution for the different confining pressure at the final slip. From top to bottom, the confining pressure is increasing. The plasticity is distributed across the sample at higher pressure while the strain is more localized in bands at lower pressures. (Simulation Parameters: $\chi_o = 0.04$, $c_o = 0.025$)

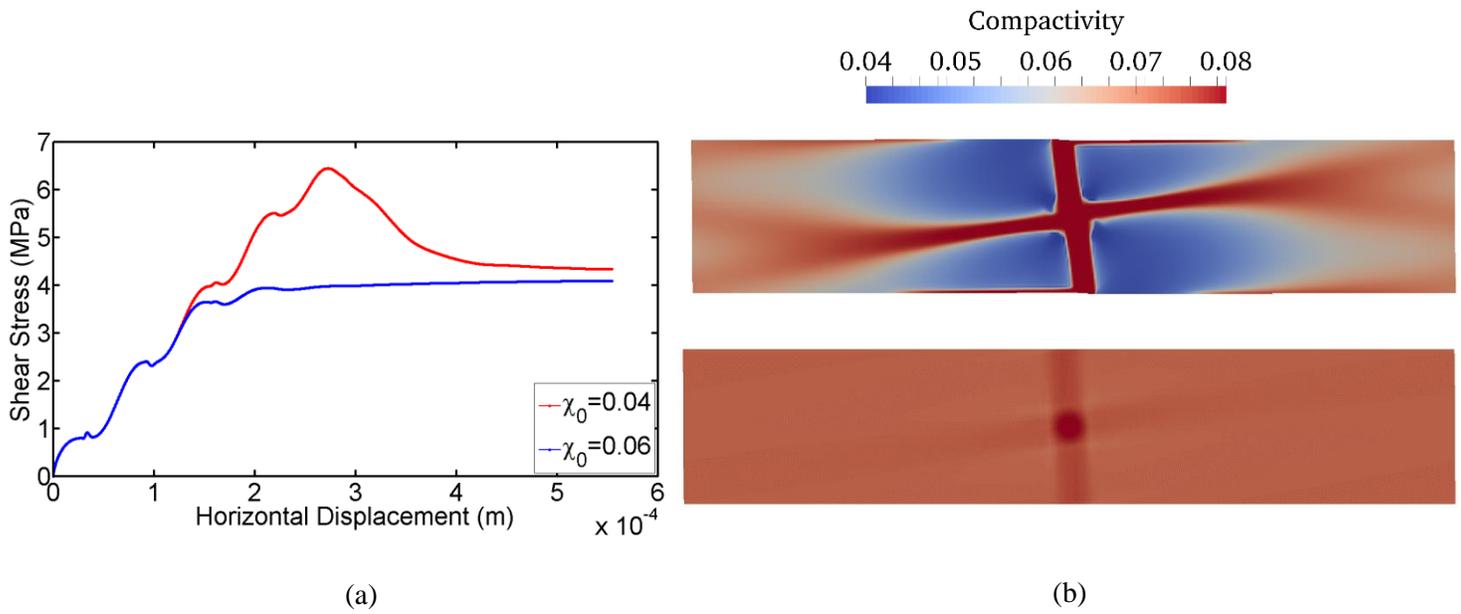

(a)                                                            (b)

**Figure 10.** Brittle to ductile transition as a function of initial compactivity. (a) Stress slip response with different initial compactivity $\chi_o$. (b) The distribution of compactivity at steady state. With lower initial compactivity, the response is brittle with strain localization softening. (Simulation Parameters: $c_o = 0.025$, $P = 10\,\text{MPa}$ )

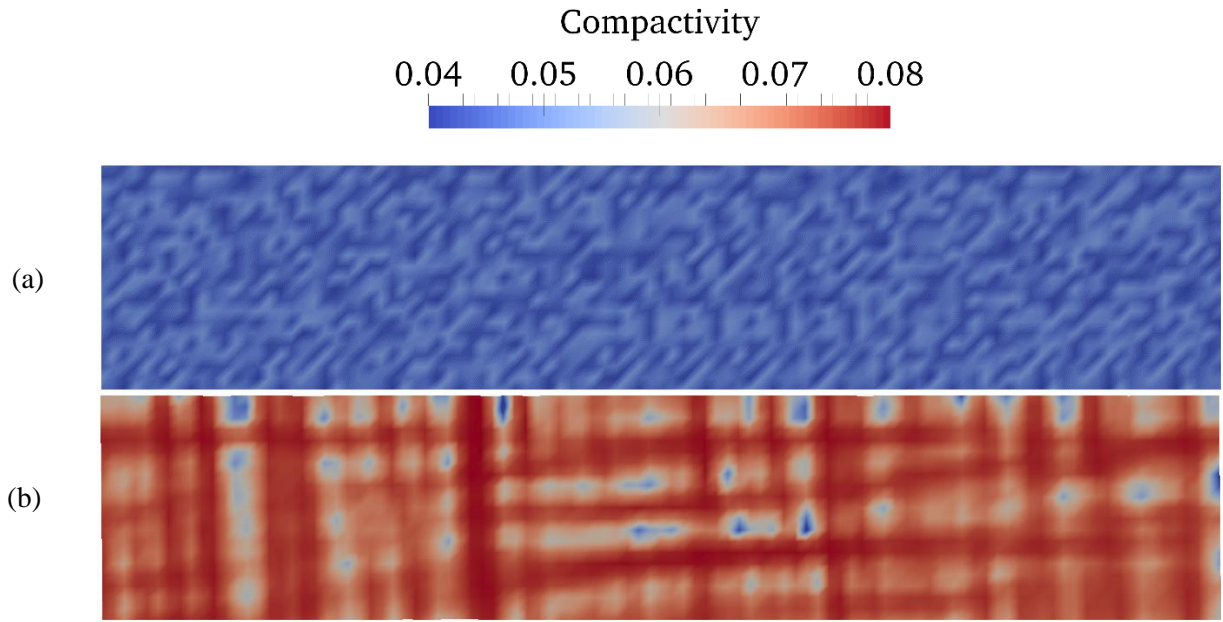

**Figure 11.** Strain localization for random initial compactivity. (a) The initial compactivity distribution is uniformly distributed in the range of 0.04~0.05. (b) The distribution of compactivity at the steady state. The random initial compactivity lead to a complex network of shear bands. (Simulation Parameters: $\chi_o = 0.04 \sim 0.05$, $c_o = 0.025$, $P = 10$ MPa)

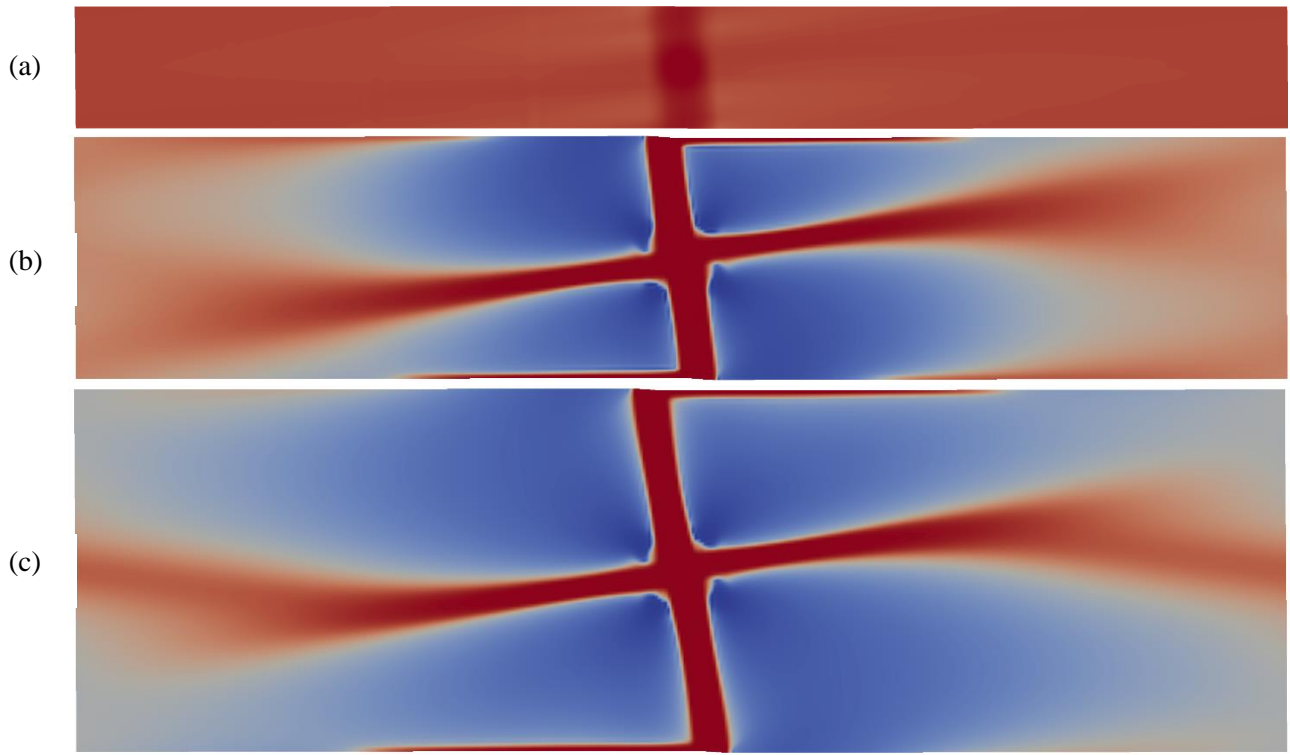

**Figure 12.** Effect of layer thickness on shear localization. (a) Thickness h = 0.1 m: the shear bands fully saturate the layer and no distinct Riedel band is formed (b) Thickness h = 0.2 m: The Riedel band forms as well as boundary shears. The Y-band starts to develop (c) Thickness h = 0.3 m: The Y-band bifurcates at an earlier stage compared to case (b) with less visible boundary shears. (Simulation Parameters: $\chi_o = 0.04$, $c_o = 0.025$, $P = 10\,\text{MPa}$)

```
For t < t_total
    At time t = t_n  χ_n, u_{x_n}, u_{y_n} are known
    Calculate  χ̇_n, the total deformation gradient F_n
    Calculate equivalent stress s̄, and the yield strength s_o
    Stress update:
    If s̄ − s_o > 0 (plasticity) then
        Formulate the residual for each plastic strain rate: R^i_{j+1} = λ̇^i_j − λ̇^i_{j+1}
        while |R| ≥ tolerance do
            Formulate Jacobian: jac = dR/dλ
            Newton-Raphson Solve for λ̇^i_{j+1}
            Update plastic deformation gradient: F^p
            Update elastic deformation gradient: F^e = F F^{p-1}
            Update Cauchy-Green Strain tensor: C^e = F^{e^T} F^e
            Update Green-Lagrangian Strain tensor: E^e = (1/2)(C^e − I)
            Update Pk2 Stress: T = L:E^e
            Update Cauchy Stress: σ = J^{e-1} F^e:T:F^{e^T}
        end do
    else (elasticity)
        Update Pk2 Stress: T = L:E^e
        Update Cauchy Stress: σ = J^{e-1} F^e:T:F^{e^T}
    end
```